# Black-Boxing the User:
# Internet Protocol over Xylophone Players (IPoXP)

**R. Stuart Geiger**
School of Information
102 South Hall
UC-Berkeley
Berkeley, CA 94720-4600
stuart@stuartgeiger.com

**Yoon Jeong**
Dept of Mechanical Engineering
2168 Etcheverry Hall
UC-Berkeley
Berkeley CA 94720
yjeong@me.berkeley.edu

**Emily Wagner**
School of Information
102 South Hall
UC-Berkeley
Berkeley, CA 94720-4600
emily@ischool.berkeley.edu

## ABSTRACT
IP over Xylophone Players (IPoXP) is an Internet connection between two computers using xylophone-based Arduino interfaces, with human operators transmitting data by striking designated keys. Inverting the traditional mode of human-computer interaction, a computer uses the human as an interface to communicate with another computer.

## ACM Classification Keywords
H.5.2 [User Interfaces]: auditory feedback, input devices and strategies, interaction styles; H.1.2 [User/Machine Systems]: human factors, models and principles; C.2.1 [Network Architecture and Design]: network communication, packet-switching networks; C.2.2 [Network Protocols]: protocol architecture (OSI model)

## INTRODUCTION
IP over Xylophone Players (IPoXP) implements a fully-compliant Internet Protocol connection between two network devices using xylophone-based Arduino interfaces, with human operators transmitting packet data by striking designated keys. At one level, IPoXP demonstrates encapsulation – the most fundamental principle of the Internet – which is that bits can be transmitted via any medium, so long as there is are standard protocols for encoding and decoding the 1s and 0s. As a standard for literally producing "tangible bits," [5] IPoXP takes the traditional infrastructure of human-computer interaction and "inverts" [9] it by situating humans as interfaces between computers. In doing so, IPoXP makes the Internet both visible and visceral, and illustrates the different levels of visibility at play in any interface, be it human-computer or computer-human. The IPoXP link is, like all networks, a socio-technical "imbroglio" [6] of human and technological actors, comprised of a temporarily stabilized assemblage of metal and plastic, operating systems and device drivers, lights and sounds, wires and electrons – as well as two human operators, who move their hands in agreement with the instructions of their network interface, striking xylophone keys when instructed in order to transmit a bit of data.

## BACKGROUND
### Protocol is as Protocol Does
At the most fundamental level, the Internet is a protocol called, simply enough, Internet Protocol (IP). As a protocol, IP is a socio-technical agreement defining how bits of information are to be circulated. IP is specifically built to be agnostic – and even blind – to how those bits are shuffled around, a principle called encapsulation. [2,4] Because of encapsulation, IP has been implemented over a number of interfaces: electrical signals through standardized wires (Ethernet, DSL, Firewire), audio links (modems), wireless radio signals (Wi-Fi, Wi-max), and infrared (IRDA). Enterprising individuals have even repurposed existing technical infrastructures and built devices which send IP traffic through domestic power lines, ham radio channels, and USB cables.

With a properly configured network interface and operating system, an application does not know – and does not need to know – the logistics of what is known as the physical layer. The web browser or chat client simply sends/receives data to/from the operating system's network stack. The network stack is also largely unaware of the specific medium used to transmit data from one source to another, delegating that task to low-level device drivers and network interface hardware such as modems. Network interfaces are typically designed for high speed, reliability, and throughput, and the dominant paradigm of network computing seeks to automate as much of these lower levels as possible.

### The OSI model
Our implementation of IPoXP can be best explained by referring to the OSI model of networking, which identifies seven different nested aspects to any communications network. Layer 1 is the physical layer, where bits are physically moved around via a communications channel, and protocols at this layer standardize electrical signals in Ethernet wires or radio channels in Wi-Fi, for example. Layer 2 is the data link layer, where the activities at the physical layer are coordinated by networking hardware, such as modems or Ethernet adapters. Layer 3 is the network layer, where the activities of the networking adapters are coordinated by a computer's operating system. Layers 4 and above define how information moves to and from different applications on a computer. All communication at layer 3 is encapsulated into IP packets, no matter which physical and data link layers are used to send and receive data.

## RELATED WORK
### Pigeons, bongo drums, and theorizing the HCI interface
The main inspiration of this project is the humorous RFC 1149, A Standard for the Transmission of IP Datagrams on Avian Carriers. [11] One of many April Fools' Day submissions, the document defines a specification for transmitting IP packets using homing pigeons. This implementation remained unused for 11 years, until 2001, when the Bergen Linux User Group set up two computers, three miles apart, each with a printer and a scanner. [1] They initiated a ping request on one computer, which printed out a sheet of paper containing a hexadecimal representation of each ICMP ping request packet. They taped the paper to the leg of a pigeon, which flew to the other site, and was removed and scanned by human operators. After scanning and OCRing the packet, a program decoded the packet and placed it in the network stack, which then delivered an ICMP ping reply packet to the printer. In all, nine packets were sent and only four were returned, with a latency of 3000-6000 seconds.

Another unconventional mode of Internet networking is Daniel Reid's The Bongo Project, [8] which used a pair of bongo drums to connect one computer to another via an IP-compliant interface. However, Reid's network interface is a fully-automated system, with a computer-controlled solenoid striking one of two bongo drums with slightly different pitches to send a 0 or 1 bit. RFC 1149 and The Bongo Project are both excellent installations for demonstrating the protocological nature of the Internet, as well as making the often-ignored bits quite visible or audible. However, both also work to reify the traditional anthropocentric mode of human-computer interaction (HCI) and computer-mediated communication (CMC), in which humans use computers to interact with other humans. The mediating term, the computer, is never simply that; rather, it is a complex assemblage of both technological and social entities that is "black boxed." [6]

What distinguishes our project from the Bergen group's implementation of RFC 1149 and Reid's The Bongo Project is that IPoXP situates humans at the lowest level of the Internet, inverting the HCI/CMC paradigm by producing an environment in which computers interact with each other via humans. In fact, to complete the experience of the human as not a user, but an interface, we literally constructed black cardboard boxes in which xylophone players would sit. With holes for two arms and a narrow window, the human operator's "Umwelt," [3,10] that is, their experiential worldview as constituted by their available sensing and acting capabilities – is reduced such that they can only see which keys light up and play the corresponding notes. Such a reversal of the roles of humans and computers inverts O'Sullivan and Igoe's famous depiction of "how the computer sees us" [7] in most desktop-based HCI designs: we appear as a giant head with two ears, but only one eye and one finger.

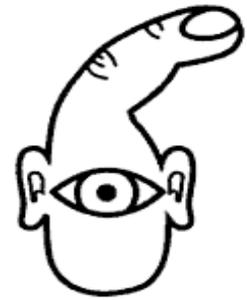

**Figure 2.** How the computer sees us, by O'Sullivan and Igoe, 2004.

## IMPLEMENTATION
### Overview
In our implementation of IPoXP, two Arduinos are connected via a USB/serial link to two laptops running a Unix-based OS, one with access to the Internet and one without. Each Arduino is connected to their local xylophone's LED ports and the remote xylophone's piezo sensors. (Figure 2)

The operation of the system as a network is best illustrated by stepping through each layer of the OSI model, introduced in the background section. As we trace the path of a packet, we show the specific role of the many different hardware, software, and human actors which make up an Internet connection. In depicting this process at the level of OSI layers, we also illustrate the quite different modes of visibility between various elements of the Internet. For example, the Umwelt of the Arduino is such that it can only interact with the computer's operating system through strings of ASCII characters; as such, it – like all network devices – is generally incapable of knowing that it is transmitting IP packets. Likewise, the Umwelt of the human operators in their confined, black boxes is such that they are largely incapable of understanding the broader significance of their tasks.

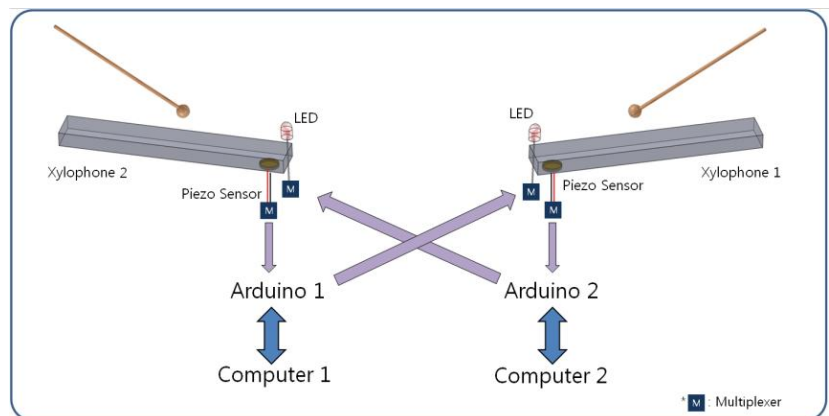

**Figure 1.** Functional diagram of the basic elements of the IPoXP interface.

**Layer 3 and above: Personal Computers**

At OSI layers 3 and above, IPoXP exists as a completely unremarkable protocol, indistinguishable to the operating system or any given application from any other network interface. Each computer establishes a Serial Line IP (SLIP) connection with a generic USB device, which from the perspective of the operating system, is an interface that sends and receives IP packets as strings of ASCII characters. When the operating system receives a command to send a 'ping' to another computer, for example, it crafts an ICMP ping request packet and translates it into ASCII. The computer then sends the ASCII-encoded packet to the generic USB device, intentionally unaware of what will happen next. If the computer receives an ASCII character from the generic USB device, it treats it as it would any piece of data transmitted over an SLIP-based connection. At layer 3, the operating system reconstructs each sequence of ASCII characters into an IP packet, even if, as in our case due to human error, the data was improperly transmitted over the physical layer.

To aid in the audience's experience of IPoXP, we created a program that was situated between layers 2 and 3 (between the Arduino's USB connection and the operating system) and would decode the constituent elements of each packet in real time. For example, when the computer received the second and third bytes of the IP packet from the Arduino, it would instantly display the packet length – which is what is to be transmitted in those two bytes (Figure 3).

**Layer 2: Arduino boards and Xylophones**

At OSI layer 2, the Arduino microprocessor board, IPoXP looks somewhat different than in layers 3 and above. The Arduino is connected to a generic serial interface over USB, which can be used to send and receive ASCII characters. The Arduino is also connected to a series of LED lights on the local xylophone keys, and a series of piezo vibration sensors on the remote xylophone's keys. When the Arduino receives a character from the computer – which is one byte or eight bits in length – it decodes the eight 1s and 0s into musical notes, and then flashes the LED corresponding to the musical note (Figure 4). When the Arduino senses that a key has been hit on the remote xylophone, it encodes the musical notes into ASCII characters, which it sends to the local computer. We must stress that aside from the human operator, each Arduino is fully independent of the other. Like with many Internet interfaces, the Arduino does not know if the other device has successfully received the packet data. If, as with some periods during our public installation, there is no human xylophone player at hand, the computer will still send packets to the Arduino, which will signal the non-present operator to hit the corresponding xylophone keys.

**Layer 1: Xylophone players**

At the lowest level of the OSI model, bits are transmitted from one Arduino to another by a human operator, who watches for the lights and hits the corresponding key on the xylophone. (Figure 5) If the xylophone player correctly enacts this process, the proper bit will be sent through the network interface – at a rate of half a baud. However, in our implementation of IPoXP, the human operator is intentionally made ignorant of this. Literally placed inside black boxes with holes for two arms and a face, the xylophone player has just enough visibility and range of movement to carry out their sole task: wait until an LED is lit, determine which xylophone key the lit LED signifies, and strike that key. Situated at OSI layer 1 of the Internet, the duties of the human operator qua network interface do not extend beyond passing bits from one location and medium to another. While we created many visualizations and feedback mechanisms for the audience to understand the process (such as a real-time description of which parts of the packet are being sent), the xylophone players remain ignorant in their black boxes.

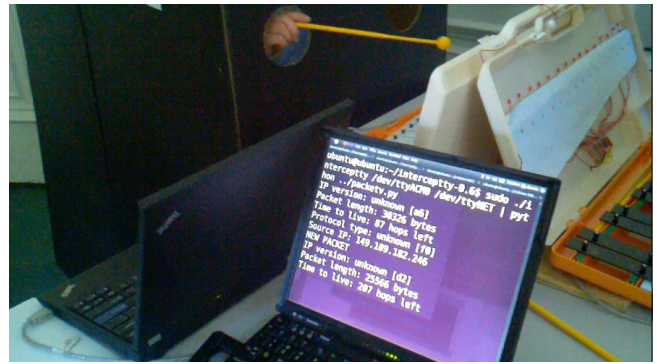

**Figure 3.** Personal computer in foreground receiving a packet from xylophone player in background

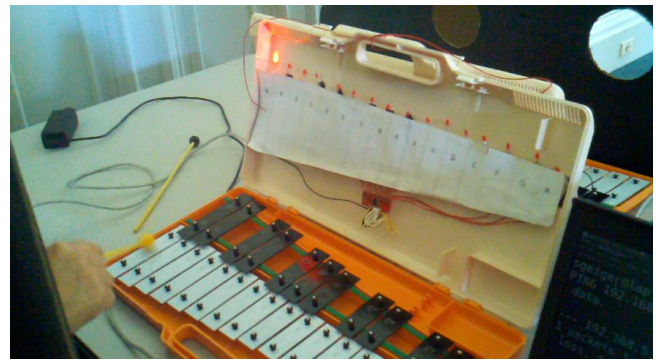

**Figure 4.** Xylophone lighting up the lower G note on the upper display, mallet striking the corresponding aluminum key

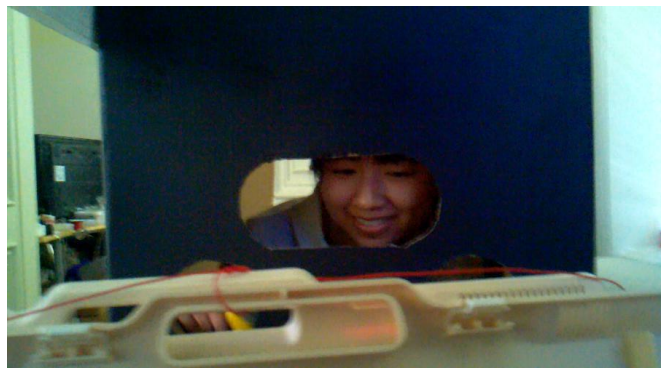

**Figure 5.** Human operator in the black box

**User and Audience Experience**
Conceptually, we wanted to push the inverted human-computer interaction paradigm as far as possible and so decided to literally "black box" the human operators, thereby abstracting the visual/spatial and neuromuscular complexity of the human brain and musculature. The black boxes were constructed out of 24" x 21" x 48" (14.0 cu/ft) cardboard boxes from our friendly neighborhood U-Haul and painted with black latex spray paint. Although the human operators inside the boxes had very little awareness of anything beyond their xylophones, we also implemented an animation using the programming language Processing which projected animated bits traveling between the two clients to emphasize the movement of bits between the two computers to the onlooking audience.

**EVALUATION AND REFLECTIONS**
Our evaluation of the interface is based on feedback received during two live demo sessions at the Tangible User Interface final project showcase at the UC Berkeley School of Information. Users came from the general public as well as fellow academics and classmates from across the university. As a result, familiarity with the standard IP network connection varied widely and explaining our adaptation and enticing visitors to try the interface proved challenging. Two visiting researchers approached the installation and immediately understood the design without needing to sit inside the boxes and play the xylophones. The two researchers also demonstrated advanced technical knowledge by calculating the connection's bandwidth (0.5 baud) when discussing the installation.

Other students would sit inside and begin dutifully striking keys but would tire before the first packet had been completed keyed – a protracted task requiring 15 minutes of one's time. A woman from another department commented that the name of the project should be changed to better reflect the empathetic experience it evoked, but liked the concept overall. Members of the general public were usually intimidated and mystified by the entire apparatus, although one person questioned the practical application of the interface, indicating that he had eventually understood the consequence of inserting a slow, unreliable human at the first level of the network stack.

After the first day of demonstrations, we learned that if this installation were to make sense on its own, that it needed some permanent explanations and signs. We added IP address signs to each network interface black box and "sender" and "receiver" labels to the two computers. Labeling the parts of the Internet with familiar text went a long way in conveying that meaning of the project.

Since this project is primarily an art installation, we envision it living in a museum like the Computer History Museum in Mountain View where it can be used, appreciated, and understood by its patrons. By contrast, putting the installation in the Explanatorium would be a less than ideal home because the target Explanatorium patron would probably be too young to have the patience or desire to play the xylophone keys in the correct order instead of extemporaneously–as children do in normal play, thereby missing the point of the installation and spoiling the child's fun.

Although we cannot completely control the amount of enjoyment people might experience while playing the IPoXP installation, the intended user experience is meant to be mechanistic and austere. This effect was achieved by limiting the player's visibility and movement while using the interface. We want the player to see the packets as we imagine computers see them, as unsentimental electrical impulses. We are also hoping that the farrago of notes coming from the xylophone as it is played to add to the sense of dystopia.

**FUTURE WORK**
Feature improvements on IPoXP might include improvements and new features including:

- integrating the projected animation with the sensed keystrokes
- making purposeful decisions about what the installation would show and/or play when not actively being used by a human operator
- giving the human operators some sort of feedback when they strike an individual key
- giving the human operators different feedback when they strike an incorrect key
- exploring how the unutilized black keys of the xylophones might be codified as larger parts of the IP header so that the human operator's time could be spent on transmitting the packet payload rather than building the header

**CONCLUSION**
This paper has presented a novel interaction technique for completing an IP network connection using xylophones. It used physical and visual constraints in order to provoke thought and empathy with the work computers perform invisibly while transmitting Internet traffic. The limited vantage point of users also forced them to focus on the low-level tedium of the IP protocol instead of at the high-level application layer, rather than optimizing the human's experience in most HCI interventions.  The ability of users to play the device depended heavily on their patience and knowledge of the OSI model, to that end the addition of familiar labels helped impart the purpose of the system.  Future work could further explore ways to make visible additional invisible layers of the IP stack.

## ACKNOWLEDGEMENTS
The authors would like to thank Kimoko Ryokai, Daniela Rosner, and Niranjan Krishnamurthi for their feedback and support and the students in the Tangible User Interfaces class at the UC Berkeley School of Information for their ideas and encouragement.

## REFERENCES
1. Bergen Linux Users Group. The highly unofficial CPIP WG. 2001. http://www.blug.linux.no/rfc1149/.
2. Cerf, V. and Kahn, R. A Protocol for Packet Network Intercommunication. *IEEE Transactions on Communications 22*, 5 (1974), 637-648.
3. Emmeche, C. Does a robot have an Umwelt? Reflections on the qualitative biosemiotics of Jakob von Uexküll. *Semiotica 2001*, 134 (2001), 653-693.
4. Galloway, A.R. *Protocol: how control exists after decentralization*. MIT Press, Cambridge, Mass., 2004.
5. Ishii, H. and Ullmer, B. Tangible bits. *Proceedings of the 1997 SIGCHI conference on Human factors in computing systems (CHI '97)*, ACM Press (1997), 234-241.
6. Latour, B. *Science in Action: How to Follow Scientists and Engineers Through Society*. Harvard University Press, Cambridge  Mass., 1987.
7. O'Sullivan, D. and Igoe, T. *Physical computing: sensing and controlling the physical world with computers*. Thompson Course Technology, Boston, 2004.
8. Reid, D. The Bongo Project. 2003. http://eagle.auc.ca/~dreid/.
9. Star, S.L. and Bowker, G.C. *Sorting Things Out: Classification and Its Consequences*. MIT Press, Cambridge, Mass., 2000.
10. von Uexküll, J. An introduction to Umwelt. *Semiotica 2001*, 134 (2001), 107-110.
11. Waitzman, D. RFC 1149:  A Standard for the Transmission of IP Datagrams on Avian Carriers . *IETF Requests for Comment*, 1990. http://www.ietf.org/rfc/rfc1149.txt.